\documentstyle[12pt]{article}
\newlength{\extraspace}
\newlength{\extraspaces}
\catcode`\@=11

\def\numberbysection{\@addtoreset{equation}{section}
\def\theequation{\arabic{section}.\arabic{equation}}}

\newcommand{\be}{\begin{equation}
\addtolength{\abovedisplayskip}{\extraspaces}
\addtolength{\belowdisplayskip}{\extraspaces}
\addtolength{\abovedisplayshortskip}{\extraspace}
\addtolength{\belowdisplayshortskip}{\extraspace}}
\newcommand{\ee}{\end{equation}}
\newcommand{\ba}{\begin{eqnarray}
\addtolength{\abovedisplayskip}{\extraspaces}
\addtolength{\belowdisplayskip}{\extraspaces}
\addtolength{\abovedisplayshortskip}{\extraspace}
\addtolength{\belowdisplayshortskip}{\extraspace}}
\newcommand{\ea}{\end{eqnarray}}
\newcommand{\nonu}{\nonumber \\[.5mm]}

\newcommand{\e}{\, {\rm e}}
\newcommand{\D}{{\cal D}}

\newcommand{\ket}[1]{\left\vert {#1} \right\rangle}
\newcommand{\VEV}[1]{\left\langle {#1} \right\rangle}

\newcommand{\sst}{\scriptstyle}
\newcommand{\sss}{\scriptscriptstyle}
\newcommand{\dash}[1]{{#1}^{\sss\prime}}
\newcommand{\ra}{\rightarrow}

\newcommand{\dd}[1]{{\partial \over \partial #1}}
\newcommand{\dds}[1]{{\partial^2 \over \partial #1{}^2}}

\newcommand{\del}{\partial}

\newcommand{\ie}{{\it i.e.}\ }

\newcommand{\ghat}{\hat{g}}
\newcommand{\chihat}{\hat{\chi}}
\newcommand{\ohat}{\hat{\Omega}}
\newcommand{\bhat}{\hat{\beta}}
\newcommand{\gahat}{\hat{\gamma}}
\newcommand{\sqghat}{\sqrt{-\ghat}}
\newcommand{\qughat}{\sqrt[\sst 4]{-\ghat}}
\newcommand{\esqghat}{\sqrt{\ghat}}
\newcommand{\equghat}{\sqrt[\sst 4]{\ghat}}
\newcommand{\ep}{\epsilon}
\newcommand{\ob}[1]{{\cal{O}}_{#1}}

\def\L{\Lambda}
\def\l{\lambda}
\def\k{\kappa}

\begin{document}
\addtolength{\baselineskip}{.7mm}
\thispagestyle{empty}
%
\begin{flushright}
TIT/HEP--210 \\
April, 1993
\end{flushright}
\vspace{3mm}
\begin{center}
{\Large{\bf Macroscopic Loop Amplitudes \\[6mm]
in Two-Dimensional Dilaton Gravity}} \\[25mm]
{\sc Yoichiro Matsumura, Norisuke Sakai and Hiroshi Shirokura} \\[14mm]
{\it Department of Physics, Tokyo Institute of Technology \\[2mm]
Oh-okayama, Meguro, Tokyo 152, Japan} \\[30mm]
%
%
{\bf Abstract}\\[8mm]
{\parbox{13cm}{\hspace{5mm}
Macroscopic loop amplitudes are obtained for the dilation gravity
in two-dimensions.
The dependence on the macroscopic loop length $l$
is completely determined by using the
Wheeler-DeWitt equation in the mini-superspace approximation.
The dependence on the cosmological constant $\Lambda$ is also determined
by using the scaling argument in addition.
}}
\end{center}
\vfill
\newpage
\setcounter{section}{0}
\setcounter{equation}{0}
\numberbysection
\section {Introduction}
Initiated by the work on the black hole evaporation \cite{CGHS},
many people have recently devoted efforts to study
two-dimensional gravity interacting with a dilaton field and matter
fields  \cite{RUTS}--\cite{MASATAUC}.
If one replaces the gravitational coupling constant $G_N$ by a spacetime
dependent field, we obtain a new gravitational theory.
By employing the exponential parametrization for the spacetime dependent
field, we have an action for the dilaton field $\phi$ and the metric
$\bar g_{\mu \nu}$
\ba
 &\!\! &\!\! S_{\rm Einstein} =
{1 \over 16\pi G_N} \int d^D x \sqrt{-\bar g} \,
\left[\bar  R  - \Lambda \right] \nonu
&\!\!  &\!\! \rightarrow
S_{\rm dilaton} = {1 \over 2\pi} \int d^D x \sqrt{-\bar g} \,
\left[  \e^{-2\phi} \bar R - 2\Lambda \right],
\label{ddimdilaton}
\ea
where we have chosen not to multiply the cosmological constant
$\Lambda$ by a function of the dilaton field $\phi$ and have adjusted
the normalization of $\Lambda$ for convenience.
A function of scalar fields multiplying the
Einstein action can be absorbed into the trace part of the
metric $\bar g_{\mu \nu}$ by a local Weyl transformation if the
spacetime dimension $D$ is different from two.
In two spacetime dimensions, the dilaton field is
of special importance for one more reason:
the Einstein action without the
dilaton field is a topological invariant which is dynamically
empty.
In order to clarify the significance of the dilaton field in two
dimensions more clearly, we make a local Weyl transformation
$g_{\mu\nu} = \e^{2\phi} \bar g_{\mu\nu}$
\be
S_{\rm dilaton} = {1 \over 2\pi} \int d^2 x \sqrt{-g} \e^{-2\phi}
\left[ R + 4 g^{\mu\nu} \partial_\mu \phi \partial_\nu \phi
- 2 \Lambda \right].
\label{cghsaction}
\ee
This form of the dilaton gravity system is suggested by
string theory and has been found to possess the black hole solution
 \cite{CGHS}.
\par
Since the back reaction of matter quantum effects can be described
by a quadratic effective action, the dilaton gravity in two-dimensions
is especially suited for studying
the back reaction of the Hawking radiation and its
consequences on the black hole evaporation.
The results of many investigations suggest that the
semi-classical approximations are
inadequate to study the black hole evaporation
\cite{RUTS}--\cite{BICA}.
Therefore it is important to consider the dilaton gravity in a
fully quantum-mechanical way.
In quantizing two-dimensional gravity, we can use
methods of conformal field theory developed for string theory
\cite{DDK}--\cite{SATAFACT}.
We have already obtained the correlation functions
of local operators as a result
of the conformal field theory treatment of dilaton gravity
\cite{MASATAUC}.
 From the viewpoint of two-dimensional gravity,
these correlation functions correspond to manifolds with several
punctures.
If we extend the punctures to macroscopic loops, we obtain
wave functions of the universe, and/or topology changing amplitudes.
In the case of the ordinary two-dimensional gravity without the dilaton
field, matrix models provide precise
methods of calculation  \cite{MOORE}, \cite{DIMOPL}.
In the continuum approach, the Wheeler-DeWitt equation
gives macroscopic loop amplitudes.
It has been shown that the solutions of the Wheeler-DeWitt equation
in the mini-superspace approximation agree completely with the results
of the matrix model \cite{MOSEST}.
We do not have
a discretized approach like the matrix model in the case of the dilaton
gravity.
However, we can formulate the Wheeler-DeWitt equation in the continuum
approach for the dilaton gravity \cite{DANIELSSON}.
\par
The purpose of this paper is to study the macroscopic loop amplitudes
in the case of the dilaton gravity.
We will solve the Wheeler-DeWitt equation in the mini-superspace
approximation in the continuum approach.
We will also apply the scaling argument
following the conformal field theory approach  \cite{DDK}
in order to examine the small
length limit of the macroscopic loop amplitude.
We find that the results of the Wheeler-DeWitt equation in the
mini-superspace approximation are in agreement with the scaling
argument and that the macroscopic loop can be represented in the small
length limit by a local operator.
Moreover, the scaling argument serves to fix the power of the
cosmological constant in the normalization of the amplitudes which
is undetermined by the Wheeler-DeWitt equation.
We also use the correlation functions of local operators to
determine the dependence of the macroscopic loop amplitudes
on the momenta of other local tachyon operators.

There has been an interesting work on the Wheeler-DeWitt equation to
obtain information on the quantum behavior of the black hole
\cite{DANIELSSON}.
On the other hand, we determine the dependence of the macroscopic loop
amplitudes on the length $l$ and the cosmological constant $\Lambda$.
Therefore our results have little overlap with theirs.
While we are writing up this work, we have received three preprints
on the quantum theory of dilaton gravity.
Two of them formulate the Wheeler-DeWitt equation and obtain its
solutions without using approximations \cite{HORI}.
Although we use the mini-superspace approximation, we have obtained
explicitly the dependence on the length $l$ and
the cosmological constant $\Lambda$ as well as on momenta of other
local operators.
The other preprint \cite{HIKASA} performed a careful canonical
quantization of the dilaton gravity.
Although they considered only
the case of a particular number of matter fields, $N=24$,
they emphasized the rigor of the analysis.
Their treatment seems to be useful especially if one wants to go beyond
the mini-superspace approximation.
\par
In Sect.2, we will describe the model of the dilaton gravity.
In Sect.3, we will present and solve the Wheeler-DeWitt equation in the
mini-superspace approximation.
In Sect.4, we will present the scaling argument.
A discussion is given in Sect.5.
\par
\section{Dilaton gravity as a conformal field theory}

We first consider a classical theory of
dilaton gravity with $N$ massless matter fields $f^j$.
We shall use the Lorentzian signature spacetime instead of the
Euclidean signature in our previous paper \cite{MASATAUC}.
In addition to the action for the dilaton gravity
(\ref{cghsaction}), we obtain the classical action
\ba
S_{\rm classical} &\!\! = &\!\! S_{\rm dilaton} + S_{\rm matter}, \nonu
S_{\rm matter}    &\!\! = &\!\! -\frac{1}{8\pi}\int\!\!d^2x\sqrt{-g}
	\sum^N_{j=1}g^{\mu\nu}\del_\mu f^j\del_\nu f^j\,.
\ea

In order to define a quantum theory for this system, we shall take the
conformal field theory approach  \cite{BICA}.
The path integral measure for the matter field is defined in terms of
the metric $g_{\mu \nu}$.
We use the conformal gauge $g_{\mu \nu}=e^{2\rho}\hat{g}_{\mu \nu}$
with the Liouville field $\rho$ and the reference
metric $\hat{g}_{\mu \nu}$.
By changing the path integral measure to the translation invariant
measure using the reference metric $\hat{g}_{\mu \nu}$, we obtain
the conformal anomaly from the matter fields represented by the
Liouville action.
As for the Liouville field, an ansatz using the conformal field theory
has been successful \cite{DDK},\cite{DHKR}.
In the case of the dilaton gravity, however, it is not obvious which
metric one should use to define the path integral measure for various
fields.
By generalizing the proposals in Refs.\ \cite{STR} and \cite{BICA},
we consider to use the metric ${\rm e}^{\alpha\phi}g_{\mu\nu}$
with various $\alpha$ for the quantization of various fields.
If we denote the amount of the anomaly for the $j$-th field
as $\gamma_j$, we find the following kinetic term for the Liouville and
the dilaton field with the parameters \cite{MASATAUC},\cite{TANII}
$a=\sum\gamma_j\alpha_j$ and $b=\sum\gamma_j\alpha_j^2$
\ba
S_{\rm kin}     &\!\! = &\!\! \frac{1}{2\pi}\int\!\!d^2x\sqghat
\bigg[
	\e^{-2\phi}\,(4\ghat^{\mu\nu}\del_\mu\phi\del_\nu\phi
	-4\ghat^{\mu\nu}\del_\mu\phi\del_\nu\rho+\hat{R})
	\nonu
 &\!\!  &\!\! +\k(\ghat^{\mu\nu}\del_\mu\rho\del_\nu\rho+\hat{R}\rho)
	-a\,(2\ghat^{\mu\nu}\del_\mu\phi\del_\nu\rho+\hat{R}\phi) \nonu
	&\!\!  &\!\!
	-b\ghat^{\mu\nu}\del_\mu\phi\del_\nu\phi
	-\frac{1}{4}\sum^N_{j=1}\ghat^{\mu\nu}\del_\mu f^j\del_\nu f^j
\bigg]\,.
\label{eqn:2_0}
\ea
We leave the total amount of the anomaly $\kappa$ as a parameter
which is to be determined by the conformal invariance.
With the translation invariant path integral measure using the reference
metric $\hat g_{\mu \nu}$, the quantum effective action consists of
the kinetic term and the
cosmological term $S_{\rm cosm}$ which will be determined later
\be
S_{\rm quantum}  =  S_{\rm kin} + S_{\rm cosm}.
\label{quantumaction}
\ee

We can perform a non-linear field transformation
in order to reduce this kinetic term
(\ref{eqn:2_0}) to a free field action \cite{MASATAUC}.
 For $\k\neq 0 $ this change of variables is
\ba
\omega  &\!\! = &\!\! \e^{-\phi},\quad
\chi = -\frac{\rho}{2}-\frac{\omega^2+a\ln\omega}{2\k} =
\frac{\hat{\chi}}{4\sqrt{|\k|}}, \nonu
d\Omega &\!\! = &\!\! -\frac{1}{\k}d\omega
\sqrt{\omega^2-\k+a+\frac{a^2+b\k}{4\omega^2}} =
\frac{d\hat{\Omega}}{4\sqrt{|\k|}}.
\ea
The Liouville field $\rho$ is contained only in the field $\hat{\chi}$.
Therefore field $\hat{\chi}$ exhibits the same transformation
property as the Liouville field under conformal transformations.
We shall call $\hat \chi$ a modified Liouville field.
After this transformation
the kinetic part of the action becomes
\be
S_{\rm kin} = \frac{1}{8\pi}\int\!\!d^2x\sqghat
\left(
	\ep\,\ghat^{\mu\nu}
	\del_\mu\chihat\del_\nu\chihat-\ep\,2\sqrt{|\k|}\hat{R}\chihat
	-\ep\,\ghat^{\mu\nu}\del_\mu\ohat\del_\nu\ohat
	-\sum^N_{j=1}\ghat^{\mu\nu}\del_\mu f^j\del_\nu f^j
\right)
\label{eqn:2_1}
\ee
where $\ep = \frac{\k}{|\k|}$.
It must be noted that
the modified Liouville field $\chihat$
has negative metric in the case of $\kappa > 0$,whereas $\ohat$
representing the dilatonic degree of freedom has negative metric
in the case of $\kappa < 0$.

The case of $\k=0$ needs a separate treatment.
We can find another change of variables
in order to transform the kinetic term to a free field action
\be
\chi^\pm = Q
\left(
	-\rho-\frac{b}{2a}\phi-\frac{2a+b}{4a}\ln
	\left|
		\e^{-2\phi}+\frac{a}{2}
	\right|
\right)
\pm\frac{2}{Q}(\e^{-2\phi}-aQ).
\ee
After this change of variables, we obtain the free field action
for the case of $\kappa=0$
\ba
S_{\rm kin} &\!\! = &\!\! \frac{1}{8\pi}\int\!\!d^2x\sqghat
\Big(
-\ghat^{\mu\nu}\del_\mu\chi^+\del_\nu\chi^+ + Q\hat{R}\chi^+ \nonu
&\!\!  &\!\! +\ghat^{\mu\nu}\del_\mu\chi^-\del_\nu\chi^- - Q\hat{R}\chi^-
	-\sum^N_{j=1}\ghat^{\mu\nu}\del_\mu f^j\del_\nu f^j
\Big).
\label{eqn:2_2}
\ea
As an $O(1,1)$ transformation in $\chi^+,\chi^-$ field space can change
the parameter $Q$, the value of $Q$ itself is not essential.

We define the path integral measure for the fields by the translation
invariant measure with the reference metric $\hat g_{\mu \nu}$.
The physical result must be independent of the choice of the
reference metric $\hat g_{\mu \nu}$.
This requirement determines the parameter $\kappa$ to be
\cite{BICA}, \cite{MASATAUC}
\be
\kappa={N-24 \over 12}.
\label{kappa}
\ee

The physical states of the theory have been obtained from the
BRST cohomology \cite{MASATAUC}.
One of them is a tachyon operator with the momentum $p$.
In the case of $N \neq 24$ ($\kappa\not=0$), the tachyon operator is
given by
\ba
\ob{p} = \int\!\!d^2x\sqghat e^{\beta_{\chi}\hat{\chi}
      +\beta_{\Omega}\hat{\Omega}
      +i\sum^N_{j=1}p_j f^j} , \nonu
\frac{1}{2} \beta_{\chi}
\left(\epsilon\beta_{\chi}-2\sqrt{|\kappa|}\right)
-\frac{1}{2}\epsilon\beta_{\Omega}^2+\frac{1}{2}\sum_j p_j^2 =1 .
\ea
In the case of $N = 24$ ($\kappa=0$), the tachyon operator is given by
\ba
\ob{p} = \int\!\!d^2x\sqghat e^{\beta_{+}\chi^{+}
      +\beta_{-}\chi^{-}
      +i\sum^N_{j=1}p_j f^j} , \nonu
-\frac{1}{2} \beta_{+} \left(\beta_{+}+Q\right)
+\frac{1}{2}\beta_{-} \left(\beta_{-}-Q\right)
+\frac{1}{2}\sum_j p_j^2 =1 .
\ea
In the case of $N\not=0$, there are also oscillator excitation
states as in the usual string theory.
Moreover, there are also other special states corresponding to the
nontrivial BRST cohomology similar to the discrete states in the
case of the Liouville gravity \cite{MASATAUC}.

Next we consider the cosmological term in (\ref{quantumaction}).
The cosmological term can be uniquely determined by requiring that
it should belong to the BRST cohomology classes and that it must coincide
with the classical one in the limit of weak gravitational coupling
constant (\ie , $\e^{\phi}\ra 0$)
\be
S_{\rm cosm} =
\left\{
\begin{array}{ll}
	-\frac{\L}{\pi}\int\!\!d^2x\sqghat\e^{\frac{1}{\sqrt{|\k|}}
	(-\chihat+\ohat)} & \quad\mbox{for}\quad\k \neq 0, \\
	-\frac{\L}{\pi}\int\!\!d^2x\sqghat\e^{-\frac{1}{Q}
	(\chi^+ +\chi^-)} & \quad\mbox{for}\quad\k = 0.
\end{array}
\right.
\label{eq:cosmoconst}
\ee

\section{The Wheeler-DeWitt equation}

The method of conformal field theory and the analytic continuation
have been used successfully to compute the correlation
functions of local operators in the case of the Liouville gravity
theory \cite{GOULI}--\cite{SATAFACT}.
However, the precise evaluation of macroscopic loop amplitudes can be
done only by means of matrix models.
Similarly, we cannot evaluate the path integral of the macroscopic
loop amplitudes directly in the continuum approach for dilaton gravity.
In the case of the Liouville gravity, the Wheeler-DeWitt equation in
the mini-superspace approximation turned out to provide solutions
which agree with the matrix model results correctly \cite{MOSEST}.
In this section we shall use the Wheeler-DeWitt equation
in the mini-superspace approximation to obtain the macroscopic
loop amplitudes.

We shall first consider the case of $N\not=24$ ($\k\neq 0$).
To formulate the Wheeler-DeWitt equations, we first construct the
energy-momentum tensor $T_{\mu\nu}$ from the action (\ref{eqn:2_1})
\ba
T_{\mu\nu} &\!\! \equiv &\!\! -4\pi\frac{1}{\sqghat}
\frac{\delta S}{\delta \ghat^{\mu\nu}}
\biggr|_{\ghat_{\mu\nu}=\eta_{\mu\nu}} \nonu
 &\!\! = &\!\! -\frac{\ep}{2}\del_\mu\chihat\del_\nu\chihat
+ \frac{\ep}{4}\eta_{\mu\nu}\del_\l\chihat\del^\l\chihat
+ \ep\sqrt{|\k|}(\eta_{\mu\nu}\del_\l\del^\l\chihat
- \del_\mu\del_\nu\chihat) \nonu
 &\!\!   &\!\! + \frac{\ep}{2}\del_\mu\ohat\del_\nu\ohat
- \frac{\ep}{4}\eta_{\mu\nu}\del_\l\ohat\del^\l\ohat
- 2\L \e^{\frac{1}{\sqrt{|\k|}}(-\chihat+\ohat)}\eta_{\mu\nu}\nonu
 &\!\!   &\!\! + \sum^N_{j=1}\biggl[\frac{1}{2}\del_\mu f^j\del_\nu f^j
- \frac{1}{4}\eta_{\mu\nu}\del_\l f^j\del^\l f^j\biggr]
\,,
\ea
where we choose the flat metric $\eta_{\mu\nu}= \mbox{diag}(-1,1)$
as the reference metric $\hat g_{\mu \nu}$.
In light-cone coordinates $x^\pm = x^0\pm x^1$, the trace part
becomes
\be
T_{-+} = T_{+-} = \ep\sqrt{|\k|}\del_+\del_-\chihat
+ \L \e^{\frac{1}{\sqrt{|\k|}}(-\chihat+\ohat)} \,.
\ee
In accordance with the conformal invariance,
the equation of motion for the modified Liouville field $\chihat$
\be
\ep \del_+ \del_- \chihat + \frac{\L }{\sqrt{|\k|}}
\e^{\frac{1}{\sqrt{|\k|}}(-\chihat+\ohat)} = 0
\label{eq:chieqofmotion}
\ee
guarantees that the trace part of the energy-momentum tensor vanishes
$T_{-+}=0$.
It is noteworthy that the equation of motion
(\ref{eq:chieqofmotion}) given by the
quantum effective action (\ref{eqn:2_1})
including the conformal anomaly effect
is precisely the condition for the vanishing of the energy-momentum
tensor.

We define the canonically conjugate momenta
 $\Pi_\chi = -\frac{\ep}{4\pi}\del_0\chihat$,
$\Pi_\Omega = \frac{\ep}{4\pi}\del_0\ohat$, and
$\Pi_j = \frac{1}{4\pi}\del_0f^j$ for
$\chihat, \ohat$, and $f^j$ respectively.
We obtain $T_{--}$ in terms of these canonical variables
\ba
T_{--} &\!\! = &\!\! -\frac{\ep}{8}(4\pi\ep\Pi_\chi
+ \partial_1\chihat)^2
- \frac{\sqrt{|\k|}}{2}\ep\partial_1(4\pi\ep\Pi_\chi + \partial_1\chihat)
\nonu
 &\!\!   &\!\! + \frac{\ep}{8}(4\pi\ep\Pi_\Omega - \partial_1\ohat)^2
+ \frac{1}{8}\sum^N_{j=1}(4\pi\Pi_j - \partial_1f^j)^2 \nonu
 &\!\!   &\!\! + \L \e^{\frac{1}{\sqrt{|\k|}}(-\chihat+\ohat)}
- \frac{\k}{2}\,.
\label{eqn:3_2}
\ea
We have added the c-number term $-\frac{\k}{2}$ in the
energy-momentum tensor $T_{--}$ because of the following
consideration \cite{SEIREV}, \cite{BRCUTH}, \cite{POLCHINSKI}.
Since we are interested in the manifold with a boundary corresponding
to the macroscopic loop, we use a coordinate system appropriate to
the geometry of cylinder.
Therefore we have to supplement the energy-momentum tensor by adding
the Schwarzian derivative term arising
from the coordinate transformation from a disk to a cylinder.
In our case, the Schwarzian derivative turns out to give
$-\k/2$.
We can expand the canonical fields in terms of oscillator modes.
 For instance, $\chihat$ and its conjugate momentum $\Pi_\chi$
can be expanded as \cite{BRCUTH}
\ba
\chihat &\!\! = &\!\! \chihat_0 + i\sum_{n\neq 0}\frac{1}{n}
\bigl(\alpha_n^\chi(x^0)\e^{inx^1}
+\tilde{\alpha}_n^\chi(x^0)\e^{-inx^1}\bigr)\,, \nonu
\Pi_\chi &\!\! = &\!\! -\frac{\ep}{4\pi}\del_0\chihat
= -\frac{\ep}{4\pi}
\Bigl[2p_0^\chi + \sum_{n\neq 0}\bigl(\alpha_n^\chi(x^0)\e^{inx^1}
+ \tilde{\alpha}_n^\chi(x^0)\e^{-inx^1}\bigr)\Bigr]\,.
\ea
It is convenient to define
$p_0^\chi = \alpha_0^\chi = \tilde{\alpha}_0^\chi$.
The mode expansions for other fields can be done similarly.
The zero-th moment $L_0$ of the energy-momentum tensor $T_{--}$ is
given by
\ba
L_0 &\!\!= &\!\! \epsilon \Bigl[-\frac{1}{2}(p_0^\chi)^2
-\sum_{m=1}^{\infty} \alpha_{-m}^{\chi}\alpha_m^{\chi}
 +  \frac{1}{2}(p_0^\Omega)^2
+ \sum_{m=1}^{\infty} \alpha_{-m}^{\Omega}\alpha_m^{\Omega} \Bigr]
- \frac{\k}{2}\nonu
 &\!\! + &\!\! \frac{\L}{2\pi}\int_0^{2\pi}\!\!dx^1
\e^{\frac{1}{\sqrt{|\k|}}(-\chihat+\ohat)}
+  \frac{1}{2}\sum^N_{j=1}(p_0^j)^2
+\sum_{m=1}^{\infty} \alpha_{-m}^{j}\alpha_m^{j}\,.
\ea
Since the metric is integrated in quantum gravity,
the energy-momentum tensor has to be treated as a constraint.
The wave function $\ket{\Psi}$ representing a macroscopic loop
amplitude should satisfy the following constraint equation
which is called the Wheeler-DeWitt equation
\be
 (L_0 - 1)\ket{\Psi} = 0\,.
\label{eqn:3_1}
\ee
Here we have assumed that the ghost part is not excited at all.
Consequently the only remnant of the ghost part is the constant term
$-1$ in the Wheeler-DeWitt equation \cite{POLCHINSKI}.

In the mini-superspace approximation, we
ignore the dependence of the fields on the spatial coordinate $x^1$.
Namely we discard all the oscillator modes of these fields.
The zero-th moment of the energy-momentum tensor becomes
in the mini-superspace approximation
\ba
L_0 &\!\! = &\!\! -\frac{\ep}{2}(p_0^\chi)^2
+ \frac{\ep}{2}(p_0^\Omega)^2
+ \frac{1}{2}\sum^N_{j=1}(p_0^j)^2 - \frac{\k}{2}
+ \L\e^{\frac{1}{\sqrt{|\k|}}(-\chihat_0+\ohat_0)} \nonu
 &\!\! = &\!\! \ep
\left[
	\frac{1}{2}\dds{\chihat_0}
	-\frac{1}{2}\dds{\ohat_0}
\right]
- \frac{\k}{2}
- \frac{1}{2}\sum^N_{j=1}\dds{f_0^j}
+\L\e^{\frac{1}{\sqrt{|\k|}}(-\chihat_0+\ohat_0)}\,.
\label{eqn:3_3}
\ea
Since the energy-momentum tensor is the sum of contributions from the
matter $f^j$ and the fields $\chi$ and $\Omega$, we can consider the
wave function
in (\ref{eqn:3_1}) as a tensor product of matter part and the rest
\be
	\ket{\Psi} = \ket{\Psi}_{\chihat,\ohat} \bigotimes
	\ket{\Psi}_{\rm matter}.
\ee
If we define the eigenvalue of $L_0$ of the matter part as $\Delta_0$
\be
-\frac{1}{2}\sum^N_{j=1}\dds{f^j_0}\ket{\Psi}_{\rm matter}
= \frac{1}{2}\sum^N_{j=1}(p^j_0)^2\ket{\Psi}_{\rm matter}
\equiv\Delta_0\ket{\Psi}_{\rm matter},
\ee
we can rewrite the Wheeler-DeWitt equation into the following form
\be
\left[
	\frac{\ep}{2}
	\left(
		\dds{\chihat_0}-\dds{\ohat_0}
	\right)
	- \biggl(1+\frac{\k}{2}-\Delta_0\biggr)
	+ \L \e^{\frac{1}{\sqrt{|\k|}}(-\chihat_0+\ohat_0)}
\right]
\Psi(\chihat_0,\ohat_0) = 0.
\label{eqn:3_4}
\ee

We now solve the Wheeler-DeWitt equation (\ref{eqn:3_4}).
It is convenient to change variables
\be
X_0 = \frac{1}{\sqrt{2}}(\ohat_0-\chihat_0),\quad
Y_0 = \frac{1}{\sqrt{2}}(\ohat_0+\chihat_0).
\label{eqn:3_9}
\ee
Depending on the choice of the physical metric,
the definition of the length $l$ of the loop may not be unique.
By regarding the cosmological constant term as the square root of the
physical metric, we shall choose the following definition of $l$
in solving the Wheeler-DeWitt equation,
\be
l=\int_0^{2\pi} {dx^1 \over 2\pi} \qughat
\e^{\frac{1}{\sqrt{|\k|}}{(-\hat \chi_0+ \hat \Omega_0) \over 2}}
=\e^{\frac{1}{\sqrt{2|\k|}}X_0},
\label{eqn:3_10}
\ee
where we denote the spatial coordinate along the boundary as $x^1$.
We find the final
form of the Wheeler-DeWitt equation for $\k\neq 0$
in terms of the length $l$ and the zero mode of the $Y$ field
\be
\left[
	l\dd{l}\dd{Y_0} + \ep\sqrt{2|\k|}
	\biggl(\frac{\k}{2}+1-\Delta_0-\L l^2\biggr)
\right]
\Psi(l,Y_0) = 0\,.
\label{eqn:3_5}
\ee

The general solution of (\ref{eqn:3_5}) can be given by
a Laplace transformed form in terms of the momentum $\hat \beta_Y$
for $Y$
\ba
\Psi(l,Y_0)
&\!\! = &\!\! \int\!\!d\bhat_Y\e^{-\bhat_YY_0}
\tilde{\Psi}(l,\bhat_Y)\,, \nonu
\tilde{\Psi}(l, \bhat_Y) &\!\! = &\!\!
C(\bhat_Y)\,
l^{\frac{1}{\bhat_Y}\ep\sqrt{2|\k|}\bigl(1+\frac{\k}{2}-\Delta_0\bigr)}
\,\e^{-\frac{\ep\sqrt{2|\k|}}{2\bhat_Y}\L l^2} \,.
\label{eqn:3_6}
\ea
We have assumed that the momentum $\hat \beta_Y \not=0$.
If $\hat \beta_Y$ vanishes, we can have an exceptional solution
proportional to the delta function for the length
$\delta\bigl(l^2-\frac{1}{\L}(1+\frac{\k}{2}-\Delta_0)\bigr)$.
If we think of this exceptional solution not just an artifact,
we have a difficulty in applying the scaling argument in the next
section. We shall discuss this in Sect.5 briefly.
A peculiar feature of the solution of the Wheeler-DeWitt equation for
the dilaton gravity is the appearance of the $Y$ momentum in the
denominator of the exponent.
This is due to the different structure of the equation compared to
the Liouville gravity case \cite{MOSEST}.

Next we discuss the case of $\k=0$.
We can obtain the energy-momentum tensor from the action
(\ref{eqn:2_2}).
The equations of motion for the $\chi_{\pm}$ fields guarantee that the
trace part of the energy-momentum tensor $T_{+-}$ vanishes.
The energy-momentum tensor $T_{--}$ is given
in light-cone coordinates
\ba
T_{--} &\!\! = &\!\! \frac{1}{8}(4\pi\Pi_+ - \partial_1\chi^+)^2
	- \frac{Q}{4}\partial_1(4\pi\Pi_+ - \partial_1\chi^+) \nonu
 &\!\!   &\!\! - \frac{1}{8}(4\pi\Pi_- + \partial_1\chi^-)^2
	- \frac{Q}{4}\partial_1(4\pi\Pi_- + \partial_1\chi^-) \nonu
 &\!\!   &\!\! + \frac{1}{8}\sum^N_{j=1}(4\pi\Pi_j + \partial_1 f^j)^2
	+ \L\e^{-\frac{1}{Q}(\chi^+ + \chi^-)}\,,
\ea
where $\Pi_+ = \frac{1}{4\pi}\del_0\chi^+$,
$\Pi_- = -\frac{1}{4\pi}\del_0\chi^-$ and
$\Pi_j = \frac{1}{4\pi}\del_0f^j$ are canonically conjugate momenta for
$\chi^+, \chi^-$ and $f^j$ respectively.
Using mode expansions of the canonical fields and the mini-superspace
approximation, we find the zero-th moment of the energy-momentum tensor
$T_{--}$
\ba
L_0 &\!\! = &\!\! \frac{1}{2}(p_0^+)^2 - \frac{1}{2}(p_0^-)^2
+ \frac{1}{2}\sum^{24}_{j=1}(p^j_0)^2
+ \L\e^{-\frac{1}{Q}(\chi^+_0 + \chi^-_0)} \nonu
 &\!\! = &\!\! -\frac{1}{2}
\left(
	\dds{\chi^+_0}-\dds{\chi^-_0}
\right)
- \frac{1}{2}\sum^{24}_{j=1}\dds{f^j_0}
+ \L\e^{-\frac{1}{Q}(\chi^+_0 + \chi^-_0)}\,.
\ea
Here the notations for the zero-modes of the canonical fields are
similar to the case of $\kappa\not=0$.
By defining the length $l$ of the macroscopic loop as
\be
l = \e^{-\frac{1}{\sqrt{2}Q}\Xi^+_0},
\quad
\Xi^\pm_0 = \frac{1}{\sqrt{2}}(\chi^-_0\pm\chi^+_0),
\label{eq:lengthn24}
\ee
we obtain the Wheeler-DeWitt equation for $\k=0$ in the
mini-superspace approximation
\be
\left[
	l\dd{l}\dd{\Xi^-_0}
	+ \sqrt{2}Q\biggl(1-\Delta_0 -\L l^2 \biggr)
\right]
\Psi(l, \Xi^-_0) = 0\,.
\label{eqn:3_7}
\ee
The solution for this case is similar to the
$\kappa\not=0$ case (\ref{eqn:3_6})
\ba
\Psi(l,\Xi^-_0) &\!\! = &\!\!\int\!\!d\gahat_
- \e^{-\gahat_-\Xi^-_0}\tilde{\Psi}(l,\gahat_-)\,, \nonu
\tilde{\Psi}(l, \gahat_-) &\!\! = &\!\!
C(\gahat_-)\,
l^{\frac{1}{\gahat_-}\sqrt{2}Q(1-\Delta_0)}
\,\e^{-\frac{Q}{\sqrt{2}\gahat_-}\L l^2}\,.
\label{eqn:3_8}
\ea
We have assumed that the momentum $\hat \gamma_- \not=0$.
If $\hat \gamma_-$ vanishes, we can have an exceptional solution
proportional to the delta function for the length
$\delta\bigl(l^2-\frac{1}{\L}(1-\Delta_0)\bigr)$.

\section{The scaling argument in the dilaton gravity}

The Wheeler-Dewitt equation determines the
wave function as a function of the length $l$ of the macroscopic loop.
However, we are interested in the dependence not only on the length $l$
but also on the cosmological constant $\Lambda$.
In order to discriminate the dependence on $l$ and $\L$,
it is useful to consider the scaling argument \cite{DDK}, \cite{MOSEST}.

In order to use the scaling argument, it is more convenient to employ
the Euclidean signature metric in this section.
It is necessary to distinguish three cases with respect to the number of
matter fields $N$.

\subsection{$N\neq 0, 24$ case}

We define the field $X$ and $Y$ as linear combinations of the fields
$\hat \chi$ and $\hat \Omega$
\be
X = \frac{1}{\sqrt{2}}(\ohat-\chihat),\qquad
 Y = \frac{1}{\sqrt{2}}(\ohat+\chihat)\,.
\ee
The momenta for $X$ and $Y$ are related to those for $\hat \chi$ and
$\hat \Omega$ as
\be
\beta_X = \frac{1}{\sqrt{2}}(\beta_\Omega - \beta_\chi)\,, \qquad
\beta_Y = \frac{1}{\sqrt{2}}(\beta_\Omega + \beta_\chi)\,.
\label{eqn:4_2}
\ee
The on-shell condition for tachyon operators is given
in terms of these variables as
\[
-\ep(\,\beta_X^{(k)} + \frac{\ep}{2}\sqrt{2|\k|}\,)
(\,\beta_Y^{(k)} - \frac{\ep}{2}\sqrt{2|\k|}\,)
= 1 + \frac{\k}{2}-\Delta_0^{(k)}\,,
\]
\be
\Delta_0^{(k)} = \frac{1}{2}\sum^N_{j=1}\,
\bigl(p_j{}^{(k)}\bigr)^2\,,\qquad(k=1,\ldots,n)\,.
\label{eqn:4_3}
\ee

We consider a manifold $\Sigma$ with $h$ handles and a boundary
together with the insertion of $n-1$ tachyon operators.
The length of the boundary is specified as $l$
and the zero mode of the $Y$ and $f^j$ fields are denoted as $Y_0$
and $f^j_0$ respectively
\be
l=\int_0^{2\pi} \frac{dx^1}{2\pi}\sqrt[\sst 4]{\ghat}
\e^{\frac{1}{\sqrt{2|\k|}}X}\,.
\ee
With these boundary conditions, the macroscopic amplitude $Z$ is
given by
\ba
\lefteqn{Z[\L, l, Y_0, f^j_0; q_1,\ldots,q_{n-1}]} \nonu
 &\!\! = &\!\! \int_{l,Y_0,f^j_0}\!\!\D X\D Y\,
	\prod^N_{i=1}(\D f^i)\,
	\e^{-S[X, Y, f^i; \L]}\,
	\prod^{n-1}_{k=1}
	\int_\Sigma\!\!d^2x_k\sqrt{\ghat(x_k)}\,
	\e^{\beta_X^{(k)}X+\beta_Y^{(k)}Y
	+i\sum_jp_j^{(k)} f^j}
	\,, \nonu
 S &\!\! = &\!\! \frac{1}{8\pi}\int_\Sigma\!\!d^2x\esqghat
	\left(
		2\ep\ghat^{\mu\nu}\del_\mu X\del_\nu Y
		+ \ep\sqrt{2|\k|}\,\hat{R}(Y-X)
		+ \sum^N_{j=1}\ghat^{\mu\nu}
		\del_\mu f^j\del_\nu f^j
	\right) \nonu
   &\!\!   &\!\! + \frac{1}{4\pi}\ep\sqrt{2|\k|}
	\oint_{\del\Sigma}\!\!ds\equghat\hat{k}(Y-X)
	+ \frac{\L}{\pi}\int_\Sigma\!\!d^2x\esqghat
	\e^{\sqrt{\frac{2}{|\k|}}\,X},
\label{eqn:4_1}
\ea
where the momenta of the tachyon operators are denoted  as
$q_k\equiv(-i\beta_X^{(k)}, -i\beta_Y^{(k)},$
$p_j^{(k)}),\ (k=1,\ldots,n-1\ \mbox{and}\ j=1,\ldots, N)$.
The integration over $X$ and $Y$ has to be performed with the boundary
condition: the length $l$ and the zero mode $Y_0$ are fixed
on the boundary.
Our macroscopic loop amplitude depends on the momenta of the
$n-1$ tachyon operators,
zero mode $Y_0$ of the $Y$ field and the length $l$ of the loop.

Let us first note that the dependence on the zero mode $Y_0$ and
$f_0^j$ of the macroscopic loop amplitude (\ref{eqn:4_1}) is given by
\ba
&\!\! &\!\! Z[\L, l, Y_0, f^j_0; q_1,\ldots,q_{n-1}] \propto
\exp \bigl(-\bhat_YY_0 + i\sum_{k=1}^{n-1} p^{(k)}_j f^j_0\bigr), \nonu
&\!\! &\!\! \bhat_Y = -
\left(
	\sum^{n-1}_{k=1}\beta_Y^{(k)} - \frac{\ep}{2}\sqrt{2|\k|}(1-2h)
\right)\,.
\label{eqn:4_4}
\ea
This implies that the macroscopic loop carries the momentum
 $-\sum_{k=1}^{n-1} p^{(k)}_j$ for the
matter fields and the momentum $\hat \beta_Y$ for the
$Y$ field as dictated by the conservation law.
Here we assigned the inward convention for the momentum of the loop.

In order to apply the scaling argument, we should make a shift of
the zero modes $X_0$ and/or $Y_0$.
Any linear combination of them can be used as long as it contains
$X_0$, since such a combination does correspond to a change of scale.
It is easy to see that any linear combination of $X_0$ and $Y_0$
will give the same result because of the momentum conservation.
Let us change the scale by performing a shift of $X_0$
\be
X \ra \dash{X} = X + \sqrt{\frac{|\k|}{2}}\rho,\quad
Y : \mbox{fixed},\quad
l \ra \dash{l} = \e^{-\frac{\rho}{2}}l\,.
\ee
Applying this zero-mode shift to the path integral (\ref{eqn:4_1}),
we find that
\ba
\lefteqn{Z[\L, l, Y_0, f^j_0;q_1,\ldots,q_{n-1}]} \nonu
 &\!\! = &\!\!
{\rm exp}\Bigl[\bigl(\frac{\k}{2}(2-2h-1)
+\sqrt{\frac{|\k|}{2}}\sum^{n-1}_{k=1}\beta_X^{(k)}\bigr)
\rho\Bigr]\,Z[\e^{\rho}\L, \e^{-\frac{\rho}{2}}l,
Y_0,f_0^j;q_1,\ldots,q_{n-1}]
\nonu
&\!\! = &\!\! \L^{-\frac{\k}{2}(1-2h)
-\sqrt{\frac{|\k|}{2}}\sum^{n-1}_{k=1}
\beta_X^{(k)}}\,
\e^{-\bhat_YY_0}\,
\e^{i\sum p_j^{(k)} f^j_0}\,
 F(\sqrt{\L}\,l).
\label{eqn:4_5}
\ea
The scaling argument gives information on
powers of dimensionful quantities such as $l$ and $\Lambda$.
Since this is precisely the role played by the scaling argument,
one should expect that the arbitrary function $F$ of the dimensionless
combination $\sqrt{\L}\,l$ cannot be determined.

Additional important information can be obtained by examining the
limit of small length $l\rightarrow 0$.
In this limit, we should obtain local operators with the momentum
dictated by the conservation law.
Possible local operators corresponding to physical states
have been classified by means of the BRST cohomology \cite{MASATAUC}.
Since we have assumed the mini-superspace approximation for the
Wheeler-DeWitt equation, we correspondingly expect local
operators without oscillator excitations.
Namely a local tachyon operator should appear in the small length
limit.
A similar assumption has been used in the case of the Liouville
gravity to give the correct result in agreement with the matrix
model \cite{MOSEST}.
We expect that the same assumption should also be valid for the
$N=0$ case, since there is no physical local operator other than
the tachyon for generic values of momenta.
 For $N\not=0$, however, we should consider also other operators in the
small length limit, if we abandon the mini-superspace approximation.
In accordance with the mini-superspace approximation, we assume here
that the macroscopic loop
in the small length limit
can be replaced by a local tachyon operator
with the momentum
$q_n\equiv(-i\beta_X^{(n)}, -i\beta_Y^{(n)}, p_j^{(n)}),\,$
$(j=1,\ldots, N)$.
We leave the power $k$ of the length $l$ in the coefficient
to be determined by the scaling argument
\be
Z[\L, l, Y_0, f^j_0;q_1,\ldots,q_{n-1}]
\stackrel{l\ra 0}{\sim}
l^k\e^{-\bhat_YY_0}
\e^{i\sum_{k=1}^{n-1} p_j^{(k)} f^j_0}
\VEV{\ob{q_1}\cdots\ob{q_n}}(\L)\,.
\label{eqn:4_6}
\ee
The $n$-point function
$\VEV{\ob{q_1}\cdots\ob{q_n}}$
denotes the correlation function of
local tachyon operators after the momentum conservation delta
functions are factored out
\ba
\lefteqn{\VEV{\ob{q_1}\cdots\ob{q_n}}(\L)} \nonu
 &\!\! = &\!\! \int\!\!\D X\D\tilde{Y}\,
	\prod^N_{i=1}(\D\tilde{f}^i)\,
	\e^{-\tilde{S}[X, \tilde{Y}, \tilde{f}^i; \L]}\,
	\prod^n_{k=1}
	\int_\Sigma\!\!d^2x_k\sqrt{\ghat(x_k)}\,
	\e^{\beta_X^{(k)}X+\beta_Y^{(k)}\tilde{Y}
	+i\sum_jp_j^{(k)}\tilde{f}^j}\,, \nonu
 \tilde{S} &\!\! = &\!\! \frac{1}{8\pi}\int_\Sigma\!\!d^2x\esqghat
	\biggl(2\ep\ghat^{\mu\nu}\del_\mu X\del_\nu\tilde{Y}
	+ \ep\sqrt{2|\k|}\hat{R}(\tilde{Y}-X)
	+ \sum^N_{j=1}\ghat^{\mu\nu}
	\del_\mu\tilde{f}^j\del_\nu\tilde{f}^j\biggr)
	\nonu
   &\!\!   &\!\! + \frac{\L}{\pi}\int_\Sigma\!\!d^2x\esqghat
	\e^{\sqrt{\frac{2}{|\k|}}X}.
\label{eqn:4_7}
\ea
We have implicitly used the coordinate system appropriate for the
cylinder geometry to describe the macroscopic amplitude near the
boundary.
Since the Liouville field has an anomalous transformation property,
we need to take into account the relation between the momentum
$\beta_Y$ for the disk geometry and the momentum $\beta_Y^{cyl}$
for the cylinder geometry \cite{POLCHINSKI}
\be
\beta_Y^{cyl}=\beta_Y-
{\epsilon \over 2}\sqrt{2|\kappa|}\,.
\ee
The momentum $\hat \beta_Y$ obtained in Eq.(\ref{eqn:4_4})
corresponds to the momentum on the cylinder.
The momentum $q_n=(-i\beta_X^{(n)}, -i\beta_Y^{(n)}, p_j^{(n)})$
of the local tachyon operator corresponding
to the shrunken loop is then given by the conservation law as shown in
Eq.(\ref{eqn:4_4})
\ba
 p_j^{(n)} &\!\! = &\!\! -\sum_{k=1}^{n-1} p^{(k)}_j\,, \nonu
\beta_Y^{(n)} &\!\! = &\!\! \hat \beta_Y^{(n)} +
{\epsilon \over 2}\sqrt{2|\kappa|}
=-
\left(
	\sum^{n-1}_{k=1}\beta_Y^{(k)} - \frac{\ep}{2}\sqrt{2|\k|}(2-2h)
\right)\,.
\ea
The on-shell condition for the tachyon $\ob{q_n}$ determines the
momentum $\beta_X^{(n)}$
\ba
-\ep(\,\beta_X^{(n)} + \frac{\ep}{2}\sqrt{2|\k|}\,)
(\beta_Y^{(n)} - \frac{\ep}{2}\sqrt{2|\k|})
&\!\! = &\!\! 1 + \frac{\k}{2}-\Delta_0^{(n)}\,, \nonu
\Delta_0^{(n)} = \frac{1}{2}\sum^N_{j=1}\bigl(p_j^{(n)}\bigr)^2\,.
\label{eqn:4_8}
\ea

By repeating the same scaling argument for the $n$-point function
of the local tachyon operators (\ref{eqn:4_7}),
we find the following $\L$ dependence
\ba
\VEV{\ob{q_1}\cdots\ob{q_n}}(\L) &\!\! = &\!\!
	\L^{-\frac{\k}{2}(2-2h)
	-\sqrt{\frac{|\k|}{2}}\sum^n_{k=1}\beta_X^{(k)}}
	\VEV{\ob{q_1}\cdots\ob{q_n}}(\L= 1) \nonu
	&\!\! = &\!\! \L^s\,\VEV{\ob{q_1}\cdots\ob{q_n}}(\L= 1),
\label{eqn:4_10}
\ea
where the exponent $s$ is the same as the one that appeared in
the calculation of the correlation function by the analytic
continuation \cite{MASATAUC}
\be
s  =  -\frac{\k}{2}(2-2h)
	-\sqrt{\frac{|\k|}{2}}\sum^n_{k=1}\beta_X^{(k)}\,.
\ee

Combining (\ref{eqn:4_6}) and (\ref{eqn:4_10}),
we find that the function $F$ in (\ref{eqn:4_5}) at
small $l$ is proportional to $(\sqrt{\L}\,l)^k$ and that
the exponent $k$ is given by
\ba
k &\!\! = &\!\! -\k - \sqrt{2|\k|}\beta_X^{(n)}
   =  \frac{\ep\sqrt{2|\k|}(1+\frac{\k}{2}-\Delta_0^{(n)})}
	{\beta_Y^{(n)}-\frac{\ep}{2}\sqrt{2|\k|}} \nonu
  &\!\! = &\!\! \frac{1}{\bhat_Y}\ep\sqrt{2|\k|}
	\bigl(1+\frac{\k}{2}-\Delta_0^{(n)}\bigr).
\label{eqn:4_11}
\ea
We see that our scaling argument is consistent with the solution
(\ref{eqn:3_6}) of the Wheeler-DeWitt equation
in the mini-superspace approximation.
Moreover we have determined the dependence of $l, \L$, and $Y_0$
in the macroscopic loop amplitude (\ref{eqn:4_1}) completely.

If we can write the path integral (\ref{eqn:4_1}) by means of
a macroscopic loop operator $W(l,\bhat_Y)$ similarly to the matrix
model, we can interpret our result by a small length expansion of the
macroscopic operator
\be
W(l,\bhat_Y) \stackrel{l\ra 0}{\sim} Cl^k\ob{q_n}+ \cdots\,.
\label{eqn:4_14}
\ee
where $\ob{q_n}$ is the local tachyon operator
with the appropriate momentum.

\subsection{$N=24$ case}

Without any additional assumption, we can repeat the scaling
argument of the preceding subsection to the case of $\k=0$.
After changing variables to
\be
\Xi^\pm=\frac{1}{\sqrt{2}}(\chi^-\pm\chi^+),\quad
\gamma_\pm=\frac{1}{\sqrt{2}}(\beta_-\pm\beta_+),
\ee
we can write the macroscopic loop amplitude as follows
\ba
Z[\L,l,\Xi^-_0,f^j_0;q_1,\cdots,q_{n-1}]
    &\!\!\! &\!\!\! = \int_{l,\Xi^-_0,f^j_0} \!\!\D\Xi^+\D\Xi^-
    \sum^{24}_{i=1}(\D f^i)
    \e^{-S[\Xi^+,\Xi^-,f^i;\L]} \nonu
 \times\! \prod^{n-1}_{k=1} &\!\!\! &\!\!\!
\int_\Sigma\!\!d^2x_k\sqrt{\ghat(x_k)}
    \e^{\gamma_+^{(k)}\Xi^++\gamma_-^{(k)}\Xi^-
    +i\sum_jp_j^{(k)}f^j}
    \!\!,
\ea
\ba
 S &\!\! = &\!\! \frac{1}{8\pi}\int_\Sigma\!\!d^2x\esqghat
    \biggl(-2\ghat^{\mu\nu}\del_\mu\Xi^+\del_\nu\Xi^-
    + \sqrt{2}Q\hat{R}\Xi^-
    + \sum^{24}_{j=1}\ghat^{\mu\nu}\del_\mu f^j\del_\nu f^j\biggr)
    \nonu
   &\!\!   &\!\! +\frac{\sqrt{2}Q}{4\pi}\oint_{\del\Sigma}\!\!ds
    \equghat\hat{k}\Xi^-
    + \frac{\L}{\pi}\int_\Sigma\!\!d^2x\esqghat\,
    \e^{-\frac{\sqrt{2}}{Q}\Xi^+}\,, \nonu
 l &\!\! = &\!\! \int_0^{2\pi}\frac{dx^1}{2\pi}\equghat\,
   \e^{-\frac{1}{\sqrt{2}Q}\Xi^+}\,.
\label{eqn:4_15}
\ea
The on-shell conditions of the $n-1$ tachyon operators are written
in the form
\ba
 &\!\! &\!\! \gamma_+^{(k)}\bigl(\gamma_-^{(k)}-\frac{Q}{\sqrt{2}}\bigr)
  = 1-\Delta_0^{(k)}\,, \nonu
 &\!\! &\!\! \Delta_0^{(k)} = \frac{1}{2}\sum^{24}_{j=1}(p_j^{(k)})^2\,,
 \quad (k=1,\ldots,n-1)\,.
\label{eq:onshell}
\ea

Evaluating the $\Xi^-_0$ dependence we find the momentum for the
macroscopic loop
\be
\gahat_- = -\biggl(\sum^{n-1}_{k=1}\gamma_-^{(k)}
           -\frac{Q}{\sqrt{2}}(1-2h)\biggr)\,.
\ee

By using the same argument as in the preceding subsection,
the zero-mode shift is performed only on the $\Xi^+$ field
\be
\Xi^+ \ra \dash{\Xi^+} = \Xi^+ - \frac{Q}{\sqrt{2}}\rho\,,\quad
\Xi^-:\mbox{fixed}\,,\quad
l \ra \dash{l} = \e^{-\frac{\rho}{2}}l\,.
\ee

We can perform the scaling argument as before and find
\be
Z[\L, l, \Xi_0^-, f^j_0; q_1,\ldots,q_{n-1}] \stackrel{l\ra 0}{\sim}
l^k\L^s\e^{-\gahat_-\Xi_0^-}\e^{i\sum^{n-1}_{m=1}p^{(m)}_jf_0^j}
\VEV{\ob{q_1}\cdots\ob{q_n}}(\L=1)\,,
\label{eq:scalingn24}
\ee
\ba
\gahat_- &\!\! = &\!\! -
\left(
	\sum^{n-1}_{k=1}\gamma_-^{(k)} - \frac{Q}{\sqrt{2}}(1-2h)
\right) \nonu
 &\!\! = &\!\! \gamma_-^{(n)} - \frac{Q}{\sqrt{2}}
   =   \gamma_-^{(n), \rm cyl}\,.
\ea
The momentum $\gahat_-$ is the $\gamma_-$ momentum of
the tachyon $\ob{q_n}$ in the cylindrical coordinate.
Since the local tachyon operator $\ob{q_n}$ is assumed to appear in the
small length limit $l \rightarrow 0$, it has the momentum
$q_n\equiv(-i\gamma_+^{(n)},-i\gamma_-^{(n)},p_j^{(n)})$,
$(j=1,\ldots,24)$ which is
determined uniquely by
the momentum conservations of the matter
fields and $\Xi_-$ field and
the on-shell condition for the tachyon
\be
\gamma_+^{(n)}
= \frac{1}{\gahat_-} (1-\Delta_0^{(n)})\,.
\ee
In this way we obtain the scaling exponents $k$ and $s$ in
Eq.(\ref{eq:scalingn24})
\be
k  = \sqrt{2}Q\gamma_+^{(n)}
= \frac{1}{\gahat_-}\sqrt{2}Q(1-\Delta_0^{(n)})\,.
\label{eqn:4_12}
\ee
\be
s  =  \frac{Q}{\sqrt{2}}\sum^n_{k=1}\gamma_+^{(k)}\,,
\label{eq:onshell24}
\ee
Thus the complete dependence on the length $l$
and the cosmological constant $\Lambda$ is determined in this case too.

\subsection{$N=0$ case}

 Finally we consider the case of no matter fields.
Though we included the $N=0$ case in the case of $N\not=24$ when
discussing the Wheeler-DeWitt equation,
we have two reasons to treat this case separately.
 First, we should introduce
the notion of chirality similarly to the $c=1$ Liouville gravity
theory, since
the on-shell condition for the tachyon operator is different
from $N\not=0$ cases.
Second, we can study a more detailed form of the macroscopic loop
amplitudes in this case,
since we can calculate the correlation
functions explicitly for an arbitrary number of local tachyon operators.

The on-shell condition in this case is given by Eq.(\ref{eqn:4_8})
\be
(\beta_X^{(k)}-1)(\beta_Y^{(k)}+1) = 0,\qquad(k=1,\ldots,n-1)\,.
\label{eqn:4_13}
\ee
Since the on-shell condition is factorized, the value of the momentum
$\beta_X$ does not specify the value of the other component $\beta_Y$
and vice versa.
This is in contrast to the on-shell condition in other cases
(\ref{eqn:4_3}) and (\ref{eq:onshell}).
We define the chirality of the tachyon to be positive
if $\beta_X=1$
and negative if $\beta_Y = -1$.
If both conditions are satisfied, the chirality is ill-defined.
This definition is quite similar to the chirality for the tachyon in the
case of the $c=1$ Liouville gravity.

By repeating the scaling argument, we find the behavior
of the macroscopic loop amplitude
for $N=0$ in the small length limit
as follows
\ba
Z[\L, l, Y_0 &\!\!; &\!\! q_1,\ldots,q_{n-1}]\stackrel{l\ra 0}{\sim}
\L^sl^k\e^{-\bhat_YY_0}f(q_1,\ldots,q_{n-1})\,, \nonu
\bhat_Y &\!\! = &\!\! \beta_Y^{(n)} + 1 \neq 0\,, \nonu
      k &\!\! = &\!\! -2(\beta_X^{(n)} - 1) = 0\,, \nonu
      s &\!\! = &\!\! 2-2h-\sum^n_{k=1}\beta_X^{(k)}\,.
\label{eqn:4_16}
\ea

This result means that the local tachyon operator corresponding to the
small length limit of the macroscopic loop has positive chirality.
The result (\ref{eqn:4_16}) is consistent with
the solution (\ref{eqn:3_6}) of the Wheeler-DeWitt equation
in the mini-superspace approximation.

 For the cases in which the matters exist, we can study
the scaling behavior of the macroscopic loop amplitude at small $l$, but
we can evaluate the correlation functions
explicitly only up to three local tachyon operators.
On the other hand,
for the $N=0$ case in which no matter
fields are present, we can evaluate the correlation
functions explicitly for an arbitrary number of tachyons.
Therefore we can obtain more detailed information of the macroscopic
loop amplitude such as the dependence on the momenta of the other
inserted tachyons.

There are two methods of computing the tachyon correlation functions
with no matter fields.
The first method is a direct evaluation of the path integral by means of
an analytic continuation.
In Ref.\cite{MASATAUC}, we computed the correlation functions
of a single negative chirality tachyon $\ob{q_1}$
and an arbitrary number of
positive chirality tachyons $\ob{q_k},\ (k=2,\ldots,n-1)$
 for spherical topology
\ba
Z[\L,l,Y_0;q_1,\ldots,q_n]_{h=0}
\stackrel{l\ra 0}{\sim} & \!\!\pi^{n-3}\!\! & \e^{-p_nY_0}\L^{2-n-p_1}
\frac{\Gamma(n-2+p_1)}{\Gamma(-p_1)}
\sum^n_{k=2}\Delta(1-p_k)\,, \nonu
p_k & \!\! = \!\! & \beta_\Omega^{(k)}\,,\quad (k=1,\ldots,n)\,,
\label{eqn:4_17}
\ea
where $\Delta(x)\equiv\frac{\Gamma(x)}{\Gamma(1-x)}$ and
$\beta_X^{(1)} = 1 + p_1,\quad \beta_Y^{(k)} = -1 + p_k$,\
$(k=2,\ldots,n)$.
It is difficult to obtain the correlation functions for other
chirality combinations by the analytic continuation \cite{DFKU}.

The second method treats this problem as a scattering problem
in a conformal field theory using the matrix model approach
\cite{DIMOPL}.
A generating function for tachyons with arbitrary momenta
has been constructed in this method.
Although it is originally derived for the $c=1$ Liouville theory,
the dilaton gravity for $N=0$ can be regarded as a condensation
of background tachyon in the $c=1$
Liouville gravity.
This background tachyon serves as the cosmological term
(\ref{eq:cosmoconst}) for the dilaton gravity.
We can apply this method even for general chirality configurations.
 For the case of single negative chirality tachyon, it gives
the same results as those by the analytic continuation.

\section{Discussion}

We have seen that
the solution of the Wheeler-DeWitt equation
is consistent with the result of the scaling argument.
However, the Wheeler-DeWitt equation admits an exceptional
solution if $\bhat_Y$ vanishes.
The solution gives the fixed length for the macroscopic
loop.
We consider this exceptional solution as an
artifact or a pathology of the Wheeler-DeWitt equation in the
mini-superspace approximation.
It seems unphysical that only a particular value is allowed for
the one-dimensional universe.
At the moment, however, we are not able to demonstrate that the
exceptional solution should be excluded.
One might need more information than the Wheeler-DeWitt equation.
It is desirable to clarify this point.

In defining the loop length $l$ in Eqs.(\ref{eqn:3_10}) or
(\ref{eq:lengthn24}), we have used the square root of the operator
defining the cosmological term (\ref{eq:cosmoconst}).
This definition is most convenient to solve the Wheeler-DeWitt equation.
However, we can choose other metric such as
$g_{\mu\nu}={\rm e}^{2\rho}\hat g_{\mu\nu}$ as the physical metric
to define the loop length $l$.
By changing variables to $\rho$ and
$\phi$, we can obtain solutions of the Wheeler-DeWitt
equation from our solution (\ref{eqn:3_6}) or (\ref{eqn:3_8})
in terms of any combinations of the Liouville and the dilaton fields
as a conformal factor.
Consequently, we obtain the solution as an integral over the unknown
weight $C(\hat \beta_Y)$ or $C(\hat \gamma_-)$.

We have considered  amplitudes with only one macroscopic loop.
It is possible to consider amplitudes with more than one macroscopic
loop.
As a first approximation, the Wheeler-DeWitt equation in the
mini-superspace approximation can be applied to find
the length dependence of each loop.

In quantizing the dilaton gravity, we have used the free field
representation together with the translation invariant measure of
these free fields.
In particular, we have assumed the path integral region
(values of the fields) to be infinite $(-\infty,\infty)$.
However, the transformation from the original variables $\rho ,\phi$ to
free fields $\hat{\chi},\hat{\Omega}$ is non-linear and quite complex.
The path integral region does not in general correspond to
the entire real values of the original variables.
It may be necessary to consider the restricted region of
the integration for the transformed variables \cite{HAMADA},
\cite{HIKASA}.
The change of the integration region does not affect the
Wheeler-DeWitt equation much, since it is a local equation.
However, it is difficult to compute the correlation functions of
local tachyon operators.
It is quite possible that some of our arguments as well as our results
need to be modified if we change the integration region.

We have chosen the present approach
motivated by the success of the continuum approach for the
Liouville gravity.
However, the lack of a more rigorous calculational tool like the
matrix model seems to suggest that we need to consider the canonical
quantization of the dilaton gravity step by step more carefully.
\par
\vspace{5mm}
%
%
{\large{\bf Acknowledgments}}
\par
We thank Y. Tanii and T. Uchino for the
collaboration in parts of this work, and Martin Hirsch for a careful
reading of the manuscript.
This work is supported in part by
Grant-in-Aide for Scientific Research for Priority Areas
from the Ministry of Education, Science and Culture (No. 04245211).
\vspace{5mm}
%
%

%
\end{document}